\documentclass[11pt]{article}

\usepackage{acl}
\usepackage{times}
\usepackage{latexsym}
\usepackage[T1]{fontenc}
\usepackage[utf8]{inputenc}
\usepackage{microtype}
\usepackage{inconsolata}
\usepackage{graphicx}

\usepackage{tikz}
\usepackage{amsmath}
\usepackage{amsfonts}
\usetikzlibrary{positioning, arrows.meta, fit, backgrounds, calc}
\usepackage{comment}
\usepackage{enumitem}
\usepackage[utf8]{inputenc}
\usepackage{booktabs}      
\usepackage{multirow}      
\usepackage{colortbl}      
\usepackage{xcolor}        
\usepackage{makecell}      
\usepackage{adjustbox}     
\usepackage{array}         
\usepackage{longtable}     
\usepackage{placeins}      
\usepackage{float}         

\title{FLINT: Fixed-budget Late Interaction}

\author{
  \textbf{Himank Handa}$^1$ \quad \textbf{Negar Arabzadeh}$^2$ \quad \textbf{Matei Zaharia}$^2$ \\
  $^1$University of Illinois Urbana-Champaign \\
  $^2$University of California, Berkeley \\
  \texttt{himankh2@illinois.edu}, \texttt{\{negara, matei\}@berkeley.edu}
}

\begin{document}
\maketitle
\begin{abstract}
Multi-vector retrieval models such as ColBERT often outperform single-vector models of comparable scale by matching queries and documents at the token level. However, retaining a vector for every token substantially increases storage requirements and slows retrieval compared to single-vector bi-encoders, limiting their scalability.
We introduce FLINT (Fixed-budget Late INTeraction), an architecture that uses attention-based pooling to represent each query and document with a fixed number of embedding vectors. Unlike architectures that pool only document representations while retaining token-level queries, FLINT applies pooling symmetrically to both sides, limiting document storage and the number of vector comparisons needed during retrieval.
FLINT achieves comparable ranking effectiveness to the closest baselines on BEIR and outperforms them on several benchmarks with longer and more complex queries, including BrowseComp+, CRUMB, and TREC Tip-of-the-Tongue 2025. We demonstrate that FLINT’s effectiveness–efficiency benefits extend across encoder architectures, testing it on two separate backbones (BERT and ModernBERT). On long-query corpora, FLINT delivers 1.5–2.8× faster end-to-end retrieval than ConstBERT, the closest baseline. FLINT also reduces index size by 5.3–9.3× relative to token-level late-interaction models such as ColBERTv2. 
\end{abstract}

\section{Introduction}
\label{sec:intro}

Multi-vector retrieval models such as ColBERT often outperform bi-encoders~\citep{reimers2019sentencebert} of comparable scale because they preserve token-level representations and enable fine-grained matching between query and document tokens \citep{khattab2020colbert}. However, this effectiveness comes at a substantial cost in latency and storage. While a standard bi-encoder stores a single vector for each document, ColBERT-style models store \textit{one vector for every document token}. Consequently, their index size grows with document length, substantially increasing storage requirements and retrieval-time computation, which leads to higher latency and limits scalability.

ColBERTv2 \citep{santhanam2022colbertv2} substantially reduces ColBERT's storage footprint through residual compression and improves retrieval effectiveness while retaining token-level representations. The PLAID index  \citep{santhanam2022plaid,macavaney2024plaid} further improves retrieval efficiency through centroid-based candidate filtering and pruning.
XTR \citep{lee2023xtr} scores documents using retrieved tokens and estimates missing token similarities. WARP \citep{scheerer2025warp} accelerates XTR-style retrieval by computing scores directly from compressed token embeddings.
MUVERA \citep{dhulipala2024muvera} uses fixed-dimensional encodings for single-vector MIPS search, followed by multi-vector reranking. Despite these advances, \textit{storage requirements remain large and tied to document length through token-level representations retained for ranking.}

\citet{constbert} address these efficiency limitations by proposing ConstBERT, in which they pool each document into a fixed set of 32 embedding vectors. This fixed representation substantially reduces index sizes relative to ColBERTv2 while preserving effectiveness on MS MARCO and BEIR. 
However, ConstBERT retains one embedding vector per query token, so the number of query vectors and the associated search cost still grow 
with longer and more complex queries, where latency also increases.
This motivates the question: \textit{Can we design a fixed-budget late-interaction architecture that preserves effectiveness on long, complex, out-of-distribution queries while reducing both document storage and query-side search cost?} We give a more detailed comparison with related work in Appendix~\ref{app:related}.

To retain the storage efficiency of fixed-budget representations without sacrificing effectiveness, and to further reduce retrieval latency, we introduce FLINT (Fixed-budget Late INTeraction).
\textbf{The key idea is to retain multiple representations for matching without tying each representation to a single input token.}
As illustrated in Figure~\ref{fig:perceiver}, learned latent vectors attend over the full token sequence, each aggregating information from multiple tokens into an output embedding. Thus, longer inputs can change the information captured by these embeddings without increasing their number.
The pooling module converts each variable-length token sequence into a fixed number of embedding vectors for multi-vector retrieval and replaces the single-vector mean-pooling operation commonly used in bi-encoders.
Figure~\ref{fig:arch_comparison} compares FLINT's architecture with prior models. Unlike ConstBERT, FLINT applies attention-based pooling symmetrically to both documents and queries. As a result, \textbf{FLINT's storage cost per document is independent of document length, and its fixed query-vector budget keeps late-interaction search latency nearly constant as query length increases}.

We evaluate FLINT on standard general-domain retrieval benchmarks, as well as out-of-domain benchmarks featuring substantially longer and more complex queries. On BEIR \citep{thakur2021beir}, FLINT outperforms the single-vector baseline and remains competitive with existing late-interaction models, while requiring substantially less storage than token-level retrieval. On more complex benchmarks, including CRUMB \citep{killingback2025crumb}, TREC Tip-of-the-Tongue 2025 \citep{arguello2026trectot}, and BrowseComp+ \citep{chen2025browsecompplus}, FLINT generally provides stronger retrieval effectiveness than ConstBERT. These results show that FLINT fixed-budget representations on both the query and document sides can retain strong retrieval effectiveness even for long, out-of-distribution queries.

FLINT combines these effectiveness results with {substantial savings in storage and retrieval time}. \textit{FLINT reduces index size by $5.3$--$9.3\times$ relative to token-level late-interaction models. }
Beyond reducing storage, {FLINT substantially improves retrieval-time efficiency}. Because it produces a fixed number of vectors for both documents and queries, its search cost remains nearly constant as query length increases. 
 On the evaluated long-query corpora, \textit{FLINT achieves 1.5--2.8$\times$ faster end-to-end retrieval} and $2.5$--$4.8\times$ faster search time than ConstBERT. 
 We note that FLINT is orthogonal to other existing efficiency techniques, including dimensionality reduction and PLAID indexing \citep{santhanam2022plaid}, which can provide complementary improvements in storage and retrieval speed.
We evaluate FLINT with ModernBERT-large~\citep{warner2024modernbert} as an alternative backbone and show that its effectiveness--efficiency benefits remain robust.

Our main contributions are:
\begin{itemize}[leftmargin=*, itemsep=2pt, topsep=2pt, parsep=0pt]
\item We introduce FLINT\footnote{Our code is available at \url{https://github.com/hmk55555/FLINT}, and model checkpoints at \url{https://huggingface.co/collections/himankhanda/flint}.
}, a fixed-budget late-interaction architecture that applies symmetric attention-based pooling to both document and query representations, decoupling storage and retrieval cost from input length.
\item We show that FLINT outperforms a single-vector baseline, remains competitive with existing late-interaction models on standard benchmarks and better preserves effectiveness on longer, out-of-distribution queries, while achieving lower retrieval latency than ConstBERT.
\item We demonstrate that FLINT generalizes across backbones and exposes the vector budget $V$ as a tunable effectiveness--efficiency trade-off.
\end{itemize}
\section{FLINT: Fixed-Budget Late Interaction}

\subsection{Fixed-Budget Representations}

Given a query $q$ and a document $d$, a shared transformer
backbone~\citep{vaswani2017attention}
independently encodes each input into a sequence of contextualized,
$h$-dimensional representations.
Let $X_q \in \mathbb{R}^{|q| \times h}$ and
$X_d \in \mathbb{R}^{|d| \times h}$ denote the resulting query and
document representations, respectively, where $|q|$ and $|d|$ are the
numbers of query and document tokens, and $h$ is the hidden dimensionality
of the transformer.

In standard late-interaction models such as ColBERT, all token representations are retained for scoring. Consequently, the number of stored document vectors grows with document length, while the cost of query--document matching grows with the number of query and document vectors.
To address this limitation, FLINT inserts an attention-based pooling module
between the transformer and the late-interaction scoring layer. The
module maps the variable-length query and document representations,
$X_q$ and $X_d$, to fixed-budget pooled representations
$Z_q, Z_d \in \mathbb{R}^{V \times h}$, where $V$ denotes the fixed number
of vectors allocated to each query and document. Thus, FLINT stores exactly
$V$ vectors per document or query, regardless of document length $|d|$ or query length $|q|$.

\subsection{Symmetric Attention-Based Pooling}

FLINT builds on the cross-attention pooling mechanism used in the Perceiver
Resampler architecture \citep{alayrac2022flamingo,jaegle2021perceiver}, which
we adapt to fixed-budget late-interaction retrieval. For each input side
$s \in \{q,d\}$, a fixed set of $V$ learned latent vectors
$L_s \in \mathbb{R}^{V \times h}$ attends over the corresponding backbone
representations $X_s$. The latent vectors act as attention queries, while the
token representations act as keys and values. This operation summarizes the
full variable-length token sequence into exactly $V$ output vectors.

For an input representation $X \in \mathbb{R}^{n \times h}$, where
$n \in \{|q|, |d|\}$ and learned latents
$L \in \mathbb{R}^{V \times h}$, the pooling operation is
{\small
\setlength{\abovedisplayskip}{4pt}
\setlength{\abovedisplayshortskip}{4pt}
\begin{equation}
\mathrm{Pool}(L, X) =
\mathrm{softmax}\!\left(
\frac{(L W_Q)(X W_K)^\top}{\sqrt{h_a}}
\right) X W_V ,
\end{equation}
}
where $W_Q$, $W_K$, and $W_V$ are learned
projection matrices, and $h_a=h/H$ is the per-head dimensionality for
$H$ attention heads. The module uses multi-head attention followed by an
output projection. Its output contains exactly $V$ vectors in
$\mathbb{R}^{V \times h}$, regardless of the input length $n$.

Each cross-attention output is then refined by a position-wise feed-forward
network applied identically and independently to each of the $V$ latent
vectors. The feed-forward network consists of an up-projection from $h$ to
$rh$ dimensions, a GELU nonlinearity, and a down-projection from $rh$ back
to $h$. Figure~\ref{fig:perceiver} illustrates our pooling
module
\footnote{We experimented with modules containing various depth self-attention stacks, but found those to lead to complete document vectors and score collapse, documented in the Appendix~\ref{app:effrank}.}.
The number of heads, the ratio $r$, and the projection dimension are given
in Appendix~\ref{app:setup}.

Document and query encoders share the same transformer backbone, pooling
module, and linear projection parameters. Only the learned latent sets,
$L_q$ and $L_d$, are specific to the query and document sides, respectively.
After pooling, the resulting representations $Z_q, Z_d$ are projected into the
retrieval embedding space and scored using ColBERT's standard late-interaction
operator. Let $z_i^q$ and $z_j^d$ denote the $i$-th and $j$-th rows of the
projected $Z_q$ and $Z_d$, i.e.\ the $i$-th projected query vector and the
$j$-th projected document vector, respectively. FLINT computes

{\small
\setlength{\abovedisplayskip}{4pt}
\setlength{\abovedisplayshortskip}{4pt}
\begin{equation}
S(q,d)
=
\sum_{i=1}^{V}
\max_{j \in \{1,\ldots,V\}}
\left(z_i^q\right)^\top z_j^d .
\end{equation}
}

\definecolor{flInput}{RGB}{255,205,210}   
\definecolor{flMHCA}{RGB}{255,204,188}    
\definecolor{flNorm}{RGB}{255,249,196}    
\definecolor{flLinear}{RGB}{225,190,231}  
\definecolor{flGELU}{RGB}{200,230,201}    
\definecolor{flDrop}{RGB}{187,222,251}    

\begin{figure}[t]
\centering
\vspace{-1em}
\resizebox{0.75\columnwidth}{!}{%
\begin{tikzpicture}[
  font=\sffamily\footnotesize,
  node distance=3mm,                 
  box/.style  = {rounded corners=1.4pt, draw=black!45, semithick, align=center,
                 minimum height=6.6mm, inner xsep=4pt, inner ysep=4pt},
  io/.style   = {box, fill=flInput!55, minimum height=6.5mm},
  norm/.style = {box, fill=flNorm!55, minimum width=18mm},
  attn/.style = {box, fill=flMHCA!55, minimum width=30mm, minimum height=8mm},
  lin/.style  = {box, fill=flLinear!55, minimum width=24mm, minimum height=7.5mm},
  gel/.style  = {box, fill=flGELU!55, minimum width=24mm, minimum height=7.5mm},
  drp/.style  = {box, fill=flDrop!55, minimum width=24mm, minimum height=7.5mm},
  outp/.style = {box, fill=black!6, minimum height=6.5mm},
  add/.style  = {circle, draw=black!60, semithick, inner sep=0pt, minimum size=3.8mm, fill=white},
  flow/.style = {-{Stealth[length=1.5mm]}, draw=black!70, semithick,
                 shorten >=1.3pt, shorten <=1.3pt},
  res/.style  = {-{Stealth[length=1.5mm]}, draw=black!40, semithick, dashed},
  klbl/.style = {font=\scriptsize\itshape, text=black!55},
]
\node[io]   (lat) {Learned latents\\ $L\in\mathbb{R}^{V\times h}$};
\node[io, right=10mm of lat] (tok) {Backbone tokens\\ $X\in\mathbb{R}^{n\times h}$};

\node[norm, below=5mm of lat] (ln1) {LayerNorm};
\node[norm] (tln) at (tok |- ln1) {LayerNorm};
\node[attn, below=of ln1] (mha) {Multi-Head\\ Cross-Attention};
\node[add,  below=of mha] (caa) {$+$};

\node[norm, below=of caa] (ln2) {LayerNorm};
\node[lin,  below=of ln2] (g1)  {Linear $h\!\to\!rh$};
\node[gel,  below=of g1]  (g2)  {GELU};
\node[lin,  below=of g2]  (g3)  {Linear $rh\!\to\!h$};
\node[drp,  below=of g3]  (g4)  {Dropout};
\node[add,  below=of g4]  (ffa) {$+$};

\node[outp, below=of ffa] (out) {Pooled latents $Z\in\mathbb{R}^{V\times h}$};

\begin{scope}[on background layer]
  \node[draw=blue!50, rounded corners=2pt, fill=blue!4, inner sep=1.8mm,
        fit=(ln1)(mha)] (cabox) {};
  \node[draw=orange!55, rounded corners=2pt, fill=orange!5, inner sep=1.8mm,
        fit=(ln2)(g4)] (ffbox) {};
\end{scope}

\draw[flow] (lat) -- (ln1);
\draw[flow] (tok) -- (tln);
\draw[flow] (ln1) -- (mha) node[klbl,midway,right=1pt] {$Q$};
\draw[flow] (tln) |- (mha.east) node[klbl,pos=0.8,above] {$K,V$};
\draw[flow] (mha) -- (caa);
\draw[flow] (caa) -- (ln2);
\draw[flow] (ln2) -- (g1);
\draw[flow] (g1)  -- (g2);
\draw[flow] (g2)  -- (g3);
\draw[flow] (g3)  -- (g4);
\draw[flow] (g4)  -- (ffa);
\draw[flow] (ffa) -- (out);

\draw[res] (lat.west) -- ++(-1.15,0) |- (caa.west);   
\draw[res] (caa.east) -- ++(2.35,0)  |- (ffa.east);   

\node[font=\normalsize\bfseries, text=blue!60, rotate=90, anchor=south]
      at ([xshift=-1.4mm]cabox.west) {Cross-Attention};
\node[font=\normalsize\bfseries, text=orange!70, rotate=90, anchor=south]
      at ([xshift=-1.4mm]ffbox.west) {Feed-Forward};

\path let \p1=(current bounding box.east) in (-\x1,0);

\end{tikzpicture}%
}
\setlength{\abovecaptionskip}{4pt}
\caption{FLINT pooling module.}
\label{fig:perceiver}
\end{figure}
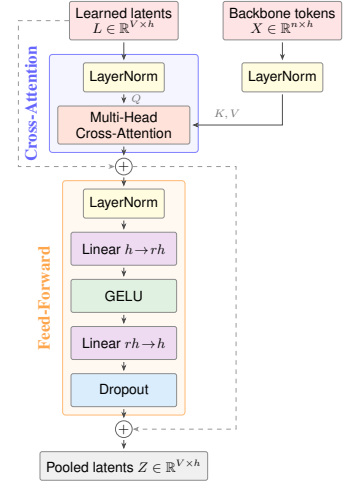

This is the same MaxSim scoring function used by ColBERTv2 and ConstBERT,
but it is computed over exactly $V$ query and $V$ document vectors rather
than over a variable number of token-level representations.
Figure~\ref{fig:arch_comparison} shows FLINT architecture relative to
baselines:
\begin{itemize}[leftmargin=*, itemsep=2pt, topsep=2pt, parsep=0pt]


\definecolor{mEnc}{RGB}{197,202,233}    
\definecolor{mAttn}{RGB}{255,204,188}   
\definecolor{mFFN}{RGB}{178,223,219}    
\definecolor{mLin}{RGB}{225,190,231}    
\definecolor{mPool}{RGB}{248,187,208}   
\definecolor{mLat}{RGB}{255,205,210}    
\definecolor{mVec}{RGB}{179,229,252}    
\definecolor{mScore}{RGB}{200,230,201}  

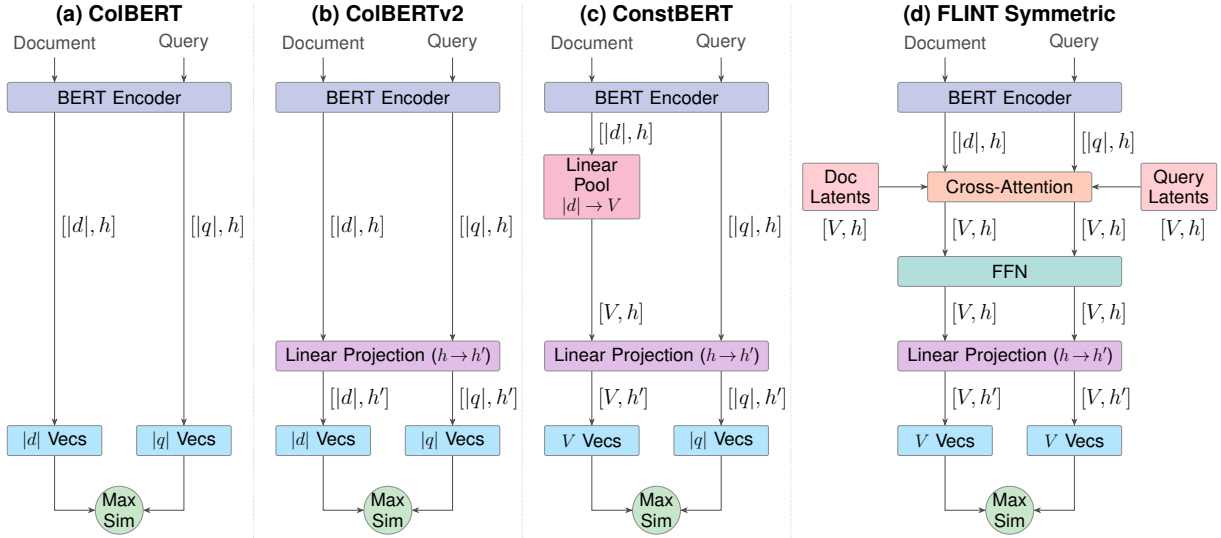
\begin{figure*}[t]
\vspace{-2em}
\centering
\resizebox{\textwidth}{!}{%
\begin{tikzpicture}[
  font=\huge\sffamily,
  mod/.style   = {rectangle, draw=black!45, semithick, align=center, rounded corners=3pt, font=\huge\sffamily},
  enc/.style   = {mod, fill=mEnc, minimum width=9.0cm, minimum height=1.2cm, text width=8.6cm},
  lin/.style   = {mod, fill=mLin, minimum width=9.0cm, minimum height=1.2cm, text width=8.6cm},
  ffn/.style   = {mod, fill=mFFN, minimum width=9.0cm, minimum height=1.2cm, text width=8.6cm},
  attn/.style  = {mod, fill=mAttn, minimum width=6.6cm, minimum height=1.2cm, text width=6.2cm},
  pool/.style  = {mod, fill=mPool, minimum width=3.8cm, minimum height=1.9cm, text width=3.6cm, rounded corners=2pt},
  latent/.style= {mod, fill=mLat, minimum width=3.0cm, minimum height=1.9cm, text width=2.9cm,
                  inner sep=2pt, rounded corners=2pt},
  vecbox/.style= {mod, fill=mVec, minimum width=3.8cm, minimum height=1.2cm, text width=3.5cm, rounded corners=2pt},
  mxs/.style   = {circle, draw=black!45, semithick, fill=mScore,
                  minimum size=1.9cm, inner sep=0pt, align=center, font=\huge\sffamily},
  arrow/.style = {-{Stealth[scale=1.3]}, semithick, draw=black!70},
  shp/.style   = {font=\fontsize{23.88}{29.86}\selectfont, text=black, inner sep=2pt, xshift=5pt},
  ttl/.style   = {font=\Huge\sffamily\bfseries},
  lbl/.style   = {font=\huge\sffamily, text=black!70},
]

\def\pz{0}   \def\pa{10.8}   \def\pb{21.6}   \def\pc{35.8}
\def\dx{2.6}   

\def\yT{18.1}    
\def\yL{17.0}    
\def\yA{16.4}    
\def\yE{14.8}    
\def\yC{11.2}    
\def\yF{7.8}     
\def\yP{4.4}     
\def\yV{1.0}     
\def\yM{-1.8}    

\node[ttl] at (\pz, \yT) {(a) ColBERT};
\node[lbl]  at (\pz-\dx, \yL) {Document};
\node[lbl]  at (\pz+\dx, \yL) {Query};

\node[enc]    (zBERT) at (\pz, \yE)  {BERT Encoder};
\node[vecbox] (zVd)   at (\pz-\dx, \yV) {$|d|$ Vecs};
\node[vecbox] (zVq)   at (\pz+\dx, \yV) {$|q|$ Vecs};
\node[mxs]    (zMs)   at (\pz, \yM) {Max\\Sim};

\draw[arrow] (\pz-\dx, \yA) -- ($(zBERT.north)+(-\dx,0)$);
\draw[arrow] (\pz+\dx, \yA) -- ($(zBERT.north)+(+\dx,0)$);

\draw[arrow] ($(zBERT.south)+(-\dx,0)$) -- (zVd.north);
\draw[arrow] ($(zBERT.south)+(+\dx,0)$) -- (zVq.north);
\node[shp, anchor=west] at (\pz-\dx, {(\yE+\yP)/2}) {$[|d|,h]$};
\node[shp, anchor=west] at (\pz+\dx, {(\yE+\yP)/2}) {$[|q|,h]$};

\draw[arrow] (zVd.south) |- (zMs.west);
\draw[arrow] (zVq.south) |- (zMs.east);

\node[ttl] at (\pa, \yT) {(b) ColBERTv2};
\node[lbl]  at (\pa-\dx, \yL) {Document};
\node[lbl]  at (\pa+\dx, \yL) {Query};

\node[enc]    (aBERT) at (\pa, \yE)  {BERT Encoder};
\node[lin]    (aLin)  at (\pa, \yP)  {Linear Projection ($h\!\to\!h'$)};
\node[vecbox] (aVd)   at (\pa-\dx, \yV) {$|d|$ Vecs};
\node[vecbox] (aVq)   at (\pa+\dx, \yV) {$|q|$ Vecs};
\node[mxs]    (aMs)   at (\pa, \yM) {Max\\Sim};

\draw[arrow] (\pa-\dx, \yA) -- ($(aBERT.north)+(-\dx,0)$);
\draw[arrow] (\pa+\dx, \yA) -- ($(aBERT.north)+(+\dx,0)$);

\draw[arrow] ($(aBERT.south)+(-\dx,0)$) -- node[right,shp]{$[|d|,h]$} ($(aLin.north)+(-\dx,0)$);
\draw[arrow] ($(aBERT.south)+(+\dx,0)$) -- node[right,shp]{$[|q|,h]$} ($(aLin.north)+(+\dx,0)$);

\draw[arrow] ($(aLin.south)+(-\dx,0)$) -- node[right,shp]{$[|d|,h']$} (aVd.north);
\draw[arrow] ($(aLin.south)+(+\dx,0)$) -- node[right,shp]{$[|q|,h']$} (aVq.north);

\draw[arrow] (aVd.south) |- (aMs.west);
\draw[arrow] (aVq.south) |- (aMs.east);

\node[ttl] at (\pb, \yT) {(c) ConstBERT};
\node[lbl]  at (\pb-\dx, \yL) {Document};
\node[lbl]  at (\pb+\dx, \yL) {Query};

\node[enc]    (bBERT) at (\pb, \yE)  {BERT Encoder};
\node[pool]   (bPool) at (\pb-\dx, \yC) {Linear Pool\\$|d|\!\to\!V$};
\node[lin]    (bLin)  at (\pb, \yP)  {Linear Projection ($h\!\to\!h'$)};
\node[vecbox] (bVd)   at (\pb-\dx, \yV) {$V$ Vecs};
\node[vecbox] (bVq)   at (\pb+\dx, \yV) {$|q|$ Vecs};
\node[mxs]    (bMs)   at (\pb, \yM) {Max\\Sim};

\draw[arrow] (\pb-\dx, \yA) -- ($(bBERT.north)+(-\dx,0)$);
\draw[arrow] (\pb+\dx, \yA) -- ($(bBERT.north)+(+\dx,0)$);

\draw[arrow] ($(bBERT.south)+(-\dx,0)$) -- node[right,shp]{$[|d|,h]$} (bPool.north);
\draw[arrow] (bPool.south) -- ($(bLin.north)+(-\dx,0)$);
\node[shp, anchor=west] at (\pb-\dx, {(\yF+\yP)/2}) {$[V,h]$};
\draw[arrow] ($(bBERT.south)+(+\dx,0)$) -- node[right,shp]{$[|q|,h]$} ($(bLin.north)+(+\dx,0)$);

\draw[arrow] ($(bLin.south)+(-\dx,0)$) -- node[right,shp]{$[V,h']$} (bVd.north);
\draw[arrow] ($(bLin.south)+(+\dx,0)$) -- node[right,shp]{$[|q|,h']$} (bVq.north);

\draw[arrow] (bVd.south) |- (bMs.west);
\draw[arrow] (bVq.south) |- (bMs.east);

\node[ttl] at (\pc, \yT) {(d) FLINT Symmetric};
\node[lbl]  at (\pc-\dx, \yL) {Document};
\node[lbl]  at (\pc+\dx, \yL) {Query};

\node[enc]    (cBERT) at (\pc, \yE)  {BERT Encoder};
\node[attn]   (cCA)   at (\pc, \yC)  {Cross-Attention};
\node[latent] (cDocLat) at (\pc-6.8, \yC) {Doc\\Latents};
\node[latent] (cQryLat) at (\pc+6.8, \yC) {Query\\Latents};
\node[ffn]    (cFFN)  at (\pc, \yF)  {FFN};
\node[shp] at ($(cDocLat |- cCA.south)!0.5!(cDocLat |- cFFN.north)$) {$[V,h]$};
\node[shp] at ($(cQryLat |- cCA.south)!0.5!(cQryLat |- cFFN.north)$) {$[V,h]$};
\node[lin]    (cLin)  at (\pc, \yP)  {Linear Projection ($h\!\to\!h'$)};
\node[vecbox] (cVd)   at (\pc-\dx, \yV) {$V$ Vecs};
\node[vecbox] (cVq)   at (\pc+\dx, \yV) {$V$ Vecs};
\node[mxs]    (cMs)   at (\pc, \yM) {Max\\Sim};

\draw[arrow] (\pc-\dx, \yA) -- ($(cBERT.north)+(-\dx,0)$);
\draw[arrow] (\pc+\dx, \yA) -- ($(cBERT.north)+(+\dx,0)$);

\draw[arrow] ($(cBERT.south)+(-\dx,0)$) -- node[right,shp]{$[|d|,h]$} ($(cCA.north)+(-\dx,0)$);
\draw[arrow] ($(cBERT.south)+(+\dx,0)$) -- node[right,shp]{$[|q|,h]$} ($(cCA.north)+(+\dx,0)$);

\draw[arrow] (cDocLat.east) -- (cCA.west);
\draw[arrow] (cQryLat.west) -- (cCA.east);

\draw[arrow] ($(cCA.south)+(-\dx,0)$) -- node[right,shp]{$[V,h]$} ($(cFFN.north)+(-\dx,0)$);
\draw[arrow] ($(cCA.south)+(+\dx,0)$) -- node[right,shp]{$[V,h]$} ($(cFFN.north)+(+\dx,0)$);

\draw[arrow] ($(cFFN.south)+(-\dx,0)$) -- node[right,shp]{$[V,h]$} ($(cLin.north)+(-\dx,0)$);
\draw[arrow] ($(cFFN.south)+(+\dx,0)$) -- node[right,shp]{$[V,h]$} ($(cLin.north)+(+\dx,0)$);

\draw[arrow] ($(cLin.south)+(-\dx,0)$) -- node[right,shp]{$[V,h']$} (cVd.north);
\draw[arrow] ($(cLin.south)+(+\dx,0)$) -- node[right,shp]{$[V,h']$} (cVq.north);

\draw[arrow] (cVd.south) |- (cMs.west);
\draw[arrow] (cVq.south) |- (cMs.east);

\draw[dotted, black!25, line width=0.9pt] (5.4,-3.0) -- (5.4, 18.6);
\draw[dotted, black!25, line width=0.9pt] (16.2,-3.0) -- (16.2, 18.6);
\draw[dotted, black!25, line width=0.9pt] (27.0,-3.0) -- (27.0, 18.6);

\end{tikzpicture}%
}
\caption{%
  {Architecture comparison across ColBERT, ColBERTv2, ConstBERT, and
  Symmetric FLINT.} Tensor shapes $[\mathrm{seq},h]$ track the sequence
  dimension through each pipeline.
}
\label{fig:arch_comparison}
\end{figure*}

\item \textbf{(a) ColBERT}~\citep{khattab2020colbert} retains one vector for
every query and document token. The resulting token-level representations are
projected and compared using MaxSim. Consequently, the document index grows
with document length, and matching cost increases with the numbers of query
and document tokens.

\item{\textbf{(b) ColBERTv2}~\citep{santhanam2022colbertv2} uses the same
token-level late-interaction architecture but applies dimensionality reduction to
reduce the storage cost of its index. However, it still retains one
representation per token, so index size and matching cost remain dependent on
document and query length. }

\item{\textbf{(c) ConstBERT}} reduces the document-side cost by pooling each document
into a fixed set of $V$ vectors ($V=32$), but it retains one representation per query
token, so retrieval cost can still increase for longer queries.

\item{\textbf{(d) FLINT}} applies the shared cross-attention and FFN pooling module
to both queries and documents. The pooling parameters are shared, while the
query and document latent vectors are learned separately. FLINT therefore
produces exactly $V$ vectors on both sides, keeping the index size fixed per
document and limiting the number of query vectors used during late-interaction
scoring.
\end{itemize}


\section{Experiments}

\subsection{Datasets}
\label{sec:datasets}

We evaluate FLINT across four complementary retrieval suites spanning diverse
tasks, queries, and corpus sizes. \textbf{BEIR}
\citep{thakur2021beir} represents conventional zero-shot retrieval across
heterogeneous domains, including question answering, fact verification, and
biomedical search, with predominantly short and keyword-like queries. \textbf{CRUMB}
\citep{killingback2025crumb}, in contrast, covers multi-aspect and
reasoning-intensive retrieval tasks whose queries express multiple  constraints, use domain-specific terminology, and often have
limited lexical overlap with relevant documents.

We additionally evaluate two retrieval settings that differ substantially from
standard passage retrieval. \textbf{TREC Tip-of-the-Tongue 2025}
\citep{arguello2026trectot} evaluates known-item re-finding from long,
conversational descriptions that may include vague details, comparisons, and
incorrect recollections.\textbf{ BrowseComp+ }\citep{chen2025browsecompplus} evaluates
deep-research retrieval, where systems must identify human-verified supporting
evidence among challenging distractor documents. Together, these suites cover
both short, conventional information needs and longer, out-of-distribution
queries. This long and complex query regime is particularly important for evaluating
FLINT because the retrieval cost of token-level late-interaction models grows
with the number of query representations. Details of  dataset statistics are
reported in Appendix~\ref{app:datasets}.

\subsection{Baselines}
\label{sec:baselines}
Our primary goal is to isolate the effectiveness and efficiency trade-offs introduced by the retrieval architecture, rather than compare independently optimized retrieval systems. We therefore focus on controlled comparisons with single-vector, token-level late-interaction, and document-only fixed-budget architectures, using matched backbones and training settings where possible.
Methods such as XTR \citep{lee2023xtr,scheerer2025warp}, token pruning \citep{acquavia2023static}, and other indexing or compression techniques address related efficiency challenges but modify different components of the retrieval pipeline. We discuss these approaches in Appendix~\ref{app:related}.
Here, we compare the results of our FLINT BERT model against ColBERTv2 \citep{santhanam2022colbertv2}
and ConstBERT \citep{constbert}. On ModernBERT, we compare against ColBERTv2, ConstBERT, and single-vector-based architectures.



\begin{table*}[t]
\vspace{-2em}
  \centering
  \footnotesize
  \captionsetup{}
  \renewcommand{\arraystretch}{1.05}

  {\footnotesize\textbf{(a)~BEIR}}\par\smallskip

  \setlength{\tabcolsep}{5pt}
  \adjustbox{max width=\linewidth}{%
  \begin{tabular}{ll|cccccccccccc|c}
    \toprule
    
    & \textbf{Architecture}
    & \shortstack{C-FEVER\\{\scriptsize $\bar q$:\,27}}
    & \shortstack{SciFact\\{\scriptsize $\bar q$:\,23}}
    & \shortstack{HotpotQA\\{\scriptsize $\bar q$:\,22}}
    & \shortstack{T-COVID\\{\scriptsize $\bar q$:\,18}}
    & \shortstack{FiQA\\{\scriptsize $\bar q$:\,16}}
    & \shortstack{SciDocs\\{\scriptsize $\bar q$:\,15}}
    & \shortstack{Quora\\{\scriptsize $\bar q$:\,14}}
    & \shortstack{FEVER\\{\scriptsize $\bar q$:\,13}}
    & \shortstack{NQ\\{\scriptsize $\bar q$:\,12}}
    & \shortstack{Touché\\{\scriptsize $\bar q$:\,10}}
    & \shortstack{DBPedia\\{\scriptsize $\bar q$:\,9}}
    & \shortstack{NFCorpus\\{\scriptsize $\bar q$:\,7}}
    & \shortstack{\textbf{Avg}\\{\scriptsize $\bar q$:\,16}} \\
    \midrule

    \multirow{3}{*}{\rotatebox[origin=c]{90}{BERT}} & FLINT ($V{=}32$)
    & 0.140 & 0.647 & 0.581 & 0.686 & 0.291 & 0.148 & 0.848 & 0.656 & 0.496 & 0.232 & 0.393 & 0.308 & 0.452 \\

     & ConstBERT
    & 0.140 & 0.618 & 0.620 & 0.738 & 0.318 & 0.159 & 0.817 & 0.683 & 0.535 & 0.253 & \underline{0.425} & 0.319 & 0.469 \\

     & ColBERTv2
    & 0.179 & 0.688 & 0.678 & 0.733 & 0.355 & 0.158 & 0.855 & 0.779 & 0.561 & 0.244 & \textbf{0.454} & 0.334 & 0.502 \\

    \midrule

     & Single-Vector
    & 0.208 & 0.604 & 0.502 & 0.794 & 0.335 & 0.156 & 0.859 & 0.621 & 0.507 & 0.224 & 0.294 & 0.285 & 0.449 \\

     & FLINT ($V{=}16$)
    & 0.215 & 0.704 & 0.612 & 0.803 & 0.380 & 0.164 & 0.872 & 0.765 & 0.555 & 0.320 & 0.396 & 0.337 & 0.510 \\

     & FLINT ($V{=}32$)
    & \textbf{0.231} & \underline{0.716} & 0.655 & \textbf{0.840} & \underline{0.403} & \underline{0.166} & \textbf{0.877} & 0.782 & 0.576 & \textbf{0.358} & 0.409 & \textbf{0.346} & 0.530 \\

     & FLINT ($V{=}64$)
    & 0.223 & \underline{0.716} & \underline{0.682} & 0.813 & \textbf{0.410} & \textbf{0.171} & 0.871 & \underline{0.797} & \underline{0.584} & 0.339 & 0.420 & \textbf{0.346} & \underline{0.531} \\

     & ConstBERT
    & \underline{0.224} & 0.702 & 0.659 & \underline{0.816} & 0.400 & \textbf{0.171} & \underline{0.875} & 0.781 & 0.551 & 0.322 & 0.378 & 0.340 & 0.518 \\

    \multirow{-6}{*}{\rotatebox[origin=c]{90}{ModernBERT}} & ColBERTv2
    & 0.203 & \textbf{0.722} & \textbf{0.720} & 0.808 & 0.397 & 0.164 & 0.869 & \textbf{0.829} & \textbf{0.592} & \underline{0.342} & 0.419 & \underline{0.343} & \textbf{0.534} \\

    \bottomrule
  \end{tabular}%
  }

  \par\vspace{0.9em}

  {\footnotesize\textbf{(b)~Complex Query Benchmarks}}\par\smallskip

  \setlength{\tabcolsep}{6pt}
  \adjustbox{max width=\linewidth}{%
  \begin{tabular}{ll|ccccccc|c|c}
    \toprule
    \multicolumn{2}{c|}{}
    & \multicolumn{7}{c|}{\textbf{CRUMB}}
    & \multicolumn{1}{c|}{\textbf{TREC ToT 2025}}
    & \multicolumn{1}{c}{\textbf{BrowseComp+}} \\
    \cmidrule(lr){3-9}
    \cmidrule(lr){10-10}
    \cmidrule(lr){11-11}

    & \textbf{Architecture}
    & \shortstack{Code\\{\scriptsize $\bar q$:\,229}}
    & \shortstack{StackEx\\{\scriptsize $\bar q$:\,226}}
    & \shortstack{Clinical\\{\scriptsize $\bar q$:\,183}}
    & \shortstack{Paper\\{\scriptsize $\bar q$:\,180}}
    & \shortstack{ToT\\{\scriptsize $\bar q$:\,168}}
    & \shortstack{Theorem\\{\scriptsize $\bar q$:\,97}}
    & \shortstack{\textbf{Avg}\\{\scriptsize $\bar q$:\,180}}
    & \shortstack{\textbf{Avg}\\{\scriptsize $\bar q$:\,204}}
    & \shortstack{\textbf{Avg}\\{\scriptsize $\bar q$:\,117}} \\
    \midrule

    \multirow{3}{*}{\rotatebox[origin=c]{90}{BERT}} & FLINT ($V{=}32$)
    & 0.045 & 0.139 & 0.264 & 0.310 & 0.078 & 0.035 & 0.145 & 0.065 & 0.063 \\

     & ConstBERT
    & 0.049 & 0.140 & 0.225 & 0.296 & 0.075 & 0.027 & 0.136 & 0.047 & 0.055 \\

     & ColBERTv2
    & 0.025 & 0.146 & 0.213 & 0.386 & 0.059 & 0.029 & 0.143 & 0.046 & 0.057 \\

    \midrule

     & Single-Vector
    & 0.055 & 0.208 & \underline{0.297} & 0.313 & 0.043 & 0.030 & 0.158 & 0.042 & 0.022 \\

     & FLINT ($V{=}16$)
    & 0.087 & \textbf{0.243} & 0.279 & 0.372 & 0.103 & 0.055 & 0.190 & 0.092 & 0.083 \\

     & FLINT ($V{=}32$)
    & \underline{0.108} & \underline{0.235} & 0.285 & \underline{0.393} & 0.138 & \textbf{0.073} & \underline{0.205} & 0.147 & 0.105 \\

     & FLINT ($V{=}64$)
    & \textbf{0.112} & 0.214 & \textbf{0.327} & \textbf{0.404} & \underline{0.184} & 0.058 & \textbf{0.217} & \textbf{0.164} & \underline{0.108} \\

     & ConstBERT
    & 0.088 & 0.207 & 0.293 & 0.330 & 0.102 & \underline{0.061} & 0.180 & 0.070 & 0.067 \\

    \multirow{-6}{*}{\rotatebox[origin=c]{90}{ModernBERT}} & ColBERTv2
    & 0.062 & 0.213 & 0.229 & 0.363 & \textbf{0.197} & 0.039 & 0.184 & \underline{0.155} & \textbf{0.135} \\

    \bottomrule
  \end{tabular}%
  }

  \caption{{Retrieval effectiveness measured using NDCG@10.} 
  \textbf{Bold} indicates the best result in each column and \underline{underline} the second best.}
  \label{tab:ndcg-results}
\end{table*}

\subsection{Training Setup \& Implementation}
\label{sec:experimental-setup}

\textbf{Training.} We train FLINT on the same MS MARCO \citep{bajaj2016msmarco} training set used in ColBERTv2 \citep{santhanam2022colbertv2} and ConstBERT \citep{constbert} so that any effectiveness gap between the three models reflects the pooling architecture rather than a difference in training data. We evaluate the trained models for ColBERT and ConstBERT on our exact setup to ensure fair comparison. In testing an alternate backbone, we replace BERT with ModernBERT while keeping FLINT's symmetric attention-based pooling architecture unchanged. Since the exact multi-vector
ModernBERT checkpoint used in prior work is not publicly available, we train the ColBERTv2 architecture on ModernBERT using the same training regimen as in \citet{warner2024modernbert}. We train FLINT under the same setup, enabling a controlled comparison between token-level and fixed-budget late interaction. We
additionally train ConstBERT on ModernBERT under the same regimen, and train FLINT with budgets $V \in \{16, 32, 64\}$; unless
noted, FLINT results use $V{=}32$. As a single-vector baseline, we use an existing ModernBERT-large bi-encoder.
Additional training, sequence lengths, query padding, and indexing details are described in Appendix~\ref{app:setup}.

\section{Retrieval Effectiveness}
\label{sec:effectiveness}
We first evaluate whether FLINT's fixed-budget representations preserve
retrieval effectiveness relative to fixed-budget and token-level
late-interaction baselines. Storage
and retrieval-time efficiency are explored separately in
Section~\ref{sec:efficiency}.

\subsection{Overall Retrieval Effectiveness}

Table~\ref{tab:ndcg-results} reports NDCG@10 on BEIR and the three
complex-query benchmarks; additional Recall@1000 results are reported separately
in Appendix Table~\ref{tab:complex-recall}. 
FLINT remains competitive with ConstBERT on BEIR, showing that
symmetric attention-based pooling preserves effectiveness on conventional,
predominantly short-query retrieval tasks. For example, on Quora, FLINT
achieves an NDCG@10 of 0.848, outperforming ConstBERT at 0.817 and closely
approaching ColBERTv2 at 0.855. ColBERTv2 nevertheless performs better on
many BEIR collections and has the highest BERT-based average, which illustrates
the effectiveness advantage that full token-level representations can retain in the short-query domain.

FLINT's advantage becomes clearer on the complex-query benchmarks. Among the
BERT-based models, FLINT achieves the highest average NDCG@10 on CRUMB,
TREC-ToT 2025, and BrowseComp+. For example, on TREC-ToT 2025, FLINT improves
NDCG@10 from 0.047 with ConstBERT to 0.065 and also exceeds ColBERTv2 at
0.046. Although the absolute scores on these benchmarks are lower than on
BEIR, this reflects the substantially greater difficulty of the tasks, which
involve long, out-of-distribution queries, limited lexical overlap, uncertain
recollections, and challenging distractor documents.

Overall, FLINT preserves competitive effectiveness on conventional retrieval
tasks while providing stronger ranking effectiveness on longer and more
complex information needs. Although token-level ColBERTv2 remains more
effective on some datasets,  FLINT's ability to retain competitive effectiveness with compact representations is valuable in light of the substantial storage and latency reductions examined in Section~\ref{sec:efficiency}.

\begin{figure*}[t]
  \centering

    \captionsetup{}
  \includegraphics[width=\textwidth]{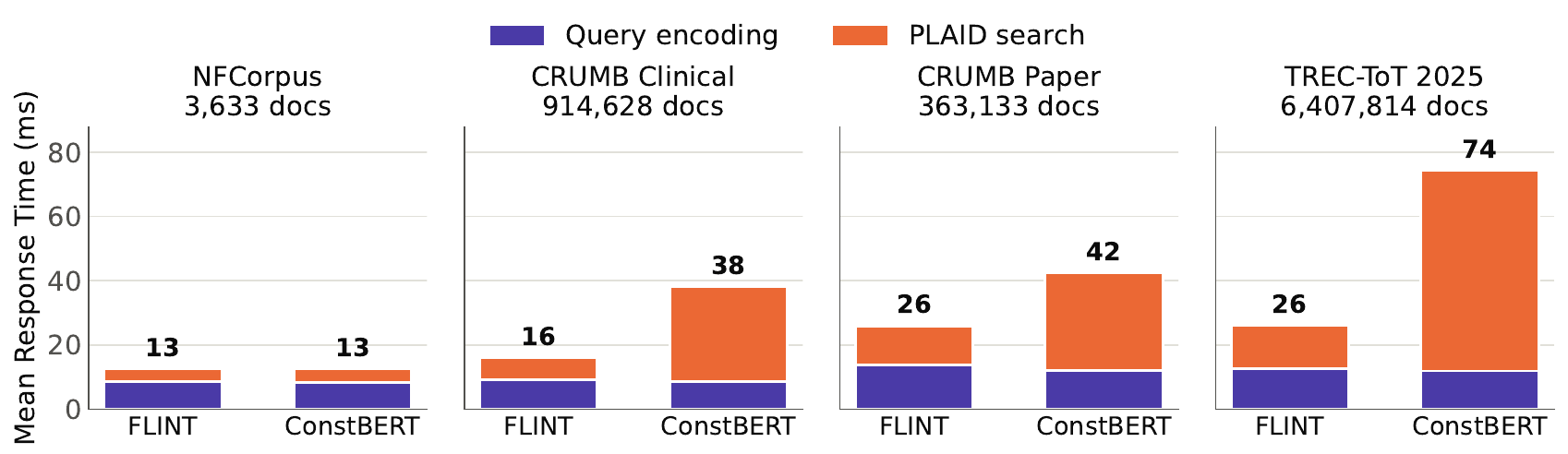}
  \caption{Mean Response Time decomposed into query-encode and
  PLAID-search time (BERT backbone), FLINT vs.\ ConstBERT, per corpus.
  Bars stack encoding (purple) and search (orange) time.}
  \label{fig:mrt-breakdown}
\end{figure*}

\begin{table*}[t]
  \centering\footnotesize
  \setlength{\tabcolsep}{3.2pt}
  \renewcommand{\arraystretch}{1.05}
  \begin{tabular}{@{}ll@{\hspace{5pt}}cc>{\columncolor{gray!12}}c@{\hspace{5pt}}cc>{\columncolor{gray!12}}c@{\hspace{5pt}}cc>{\columncolor{gray!12}}c@{\hspace{5pt}}cc>{\columncolor{gray!12}}c@{}}
  \toprule
   & & \multicolumn{3}{c}{\makecell{\textbf{NFCorpus}\\{\scriptsize 3.6K docs, $\bar q$: 7}}} & \multicolumn{3}{c}{\makecell{\textbf{CRUMB Paper}\\{\scriptsize 363K docs, $\bar q$: 180}}} & \multicolumn{3}{c}{\makecell{\textbf{CRUMB Clinical}\\{\scriptsize 915K docs, $\bar q$: 183}}} & \multicolumn{3}{c}{\makecell{\textbf{TREC-ToT 2025}\\{\scriptsize 6.4M docs, $\bar q$: 204}}} \\
  \cmidrule(lr){3-5}\cmidrule(lr){6-8}\cmidrule(lr){9-11}\cmidrule(l){12-14}
   & \textbf{Architecture} & \textbf{Enc.} & \textbf{Srch.} & \textbf{MRT}\,$\downarrow$ & \textbf{Enc.} & \textbf{Srch.} & \textbf{MRT}\,$\downarrow$ & \textbf{Enc.} & \textbf{Srch.} & \textbf{MRT}\,$\downarrow$ & \textbf{Enc.} & \textbf{Srch.} & \textbf{MRT}\,$\downarrow$ \\
  \midrule
  \multirow{3}{*}{\rotatebox[origin=c]{90}{BERT}} & FLINT ($V{=}32$) & 8.5 & 4.1 & \textbf{12.6} & 13.8 & 12.1 & \textbf{25.9} & 9.2 & 6.9 & \textbf{16.1} & 12.6 & 13.5 & \textbf{26.1} \\
   & ConstBERT & 8.3 & 4.3 & \textbf{12.6} & 12.1 & 30.3 & \underline{42.4} & 8.7 & 29.5 & \underline{38.2} & 11.9 & 62.4 & \underline{74.3} \\
   & ColBERTv2 & 8.4 & 6.8 & \underline{15.2} & 11.8 & 259.0 & 270.8 & 8.4 & 574.3 & 582.7 & 11.8 & 763.1 & 774.9 \\
  \midrule
  \multirow{6}{*}{\rotatebox[origin=c]{90}{ModernBERT}} & Single-Vector & 16.9 & 1.0 & \textbf{17.9} & 22.0 & 1.1 & \textbf{23.1} & 17.2 & 1.0 & \textbf{18.2} & 22.7 & 1.4 & \textbf{24.1} \\
   & FLINT ($V{=}16$) & 19.1 & 4.4 & 23.5 & 26.0 & 7.7 & \underline{33.7} & 20.5 & 6.7 & \underline{27.2} & 27.6 & 12.1 & \underline{39.7} \\
   & FLINT ($V{=}32$) & 19.1 & 4.6 & 23.7 & 26.6 & 11.3 & 37.9 & 20.4 & 9.3 & 29.7 & 27.7 & 23.6 & 51.3 \\
   & FLINT ($V{=}64$) & 18.9 & 5.5 & 24.4 & 26.7 & 31.4 & 58.1 & 20.6 & 45.2 & 65.8 & 27.3 & 75.1 & 102.4 \\
   & ConstBERT & 18.5 & 4.3 & \underline{22.8} & 27.1 & 28.3 & 55.4 & 20.1 & 34.7 & 54.8 & 26.8 & 114.1 & 140.9 \\
   & ColBERTv2 & 16.0 & 10.4 & 26.4 & 21.7 & 842.0 & 863.7 & 16.4 & 2118.1 & 2134.5 & 22.1 & 8685.8 & 8707.9 \\
  \bottomrule
  \end{tabular}
  \caption{Mean response time (ms). Query encoding (Enc.) plus search (Srch.) gives the
  mean response time (MRT). For each backbone and corpus, \textbf{bold} marks the lowest
  MRT and \underline{underline} the second lowest. $\bar q$ is the mean query length in
  tokens, and $V$ is the number of vectors per query and document.}
  \label{tab:mrt}
\end{table*}

\subsection{Impact of the Backbone Encoder}
\label{sec:modernbert-backbone}

We next examine how the backbone encoder affects FLINT's retrieval
effectiveness.

As shown in Table~\ref{tab:ndcg-results}, replacing BERT with ModernBERT
consistently improves FLINT across all benchmark suites. Average NDCG@10
increases from 0.452 to 0.530 on BEIR and from 0.145 to 0.205 on CRUMB.
The gains are particularly pronounced on the complex-query benchmarks. 
FLINT improves from 0.065 to 0.147 on TREC-ToT 2025 and from 0.063 to
0.105 on BrowseComp+. These results show that FLINT benefits substantially
from a stronger backbone, especially on longer and out-of-distribution
information needs.

With the backbone and training held fixed, FLINT outperforms ConstBERT on
every suite on ModernBERT. The gap is largest on TREC-ToT 2025, whose queries are the longest
and most conversational, where ConstBERT reaches less than half of FLINT's
score. 
Compared with ModernBERT ColBERTv2, FLINT is nearly tied on
BEIR (0.530 vs.\ 0.534) and outperforms it on CRUMB (0.205 vs.\ 0.184). ModernBERT ColBERTv2 remains ahead on TREC-ToT 2025 and BrowseComp+. FLINT is stronger than the
single-vector baseline, particularly on long queries, where NDCG@10 rises
from 0.042 to 0.147 on TREC-ToT 2025 and from 0.022 to 0.105 on BrowseComp+. This indicates that retaining multiple pooled
vectors, rather than pooling to a single vector, is central to its
effectiveness.

\subsection{Impact of the Budget $V$}
\label{sec:budget}
We examine how the number of vectors per query and document, $V$, affects FLINT's effectiveness with the ModernBERT backbone. Reducing the budget from $V{=}32$ to $V{=}16$ lowers effectiveness across all benchmark suites. Nevertheless, FLINT with $V{=}16$ still outperforms the single-vector baseline on every suite and ModernBERT ConstBERT on all three complex-query suites, while requiring approximately half of ConstBERT's index size.

Increasing the budget from $V{=}32$ to $V{=}64$ leaves average BEIR effectiveness nearly unchanged (0.531 vs.\ 0.530) but improves performance on complex queries. 
These results suggest that longer, multi-constraint queries benefit from a larger vector budget, whereas the predominantly short queries in BEIR gain little beyond $V{=}32$. \textbf{The budget thus provides a direct way to trade additional storage and computation for improved effectiveness on complex queries.}

\begin{figure*}[h]
  \centering
  \vspace{-1em}
  \captionsetup{font=footnotesize}
  \includegraphics[width=\textwidth]{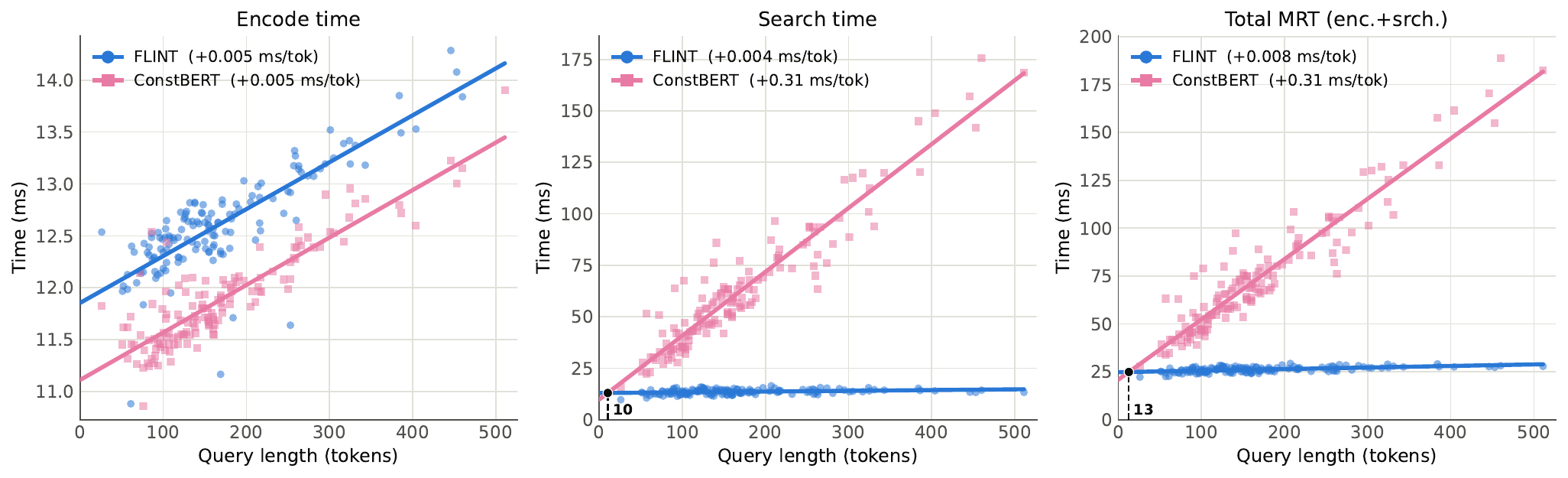}
  \caption{Response-time decomposition on TREC-ToT~2025 (BERT backbone).
  Query-encode (left), PLAID-search (middle), and total MRT (right) against
  query length for FLINT and ConstBERT, with linear fits (slopes in the legend);
  dashed drops mark where the fits cross.}
  \label{fig:mrt-decompose}
\end{figure*}

\section{Efficiency Analysis}
\label{sec:efficiency}

We next evaluate FLINT's storage and retrieval-time efficiency relative to baselines on NFCorpus, CRUMB Paper, CRUMB Clinical, and TREC-ToT 2025. We select these corpora to examine how efficiency varies with both corpus size and query length. They span 3.6K to 6.4M documents and mean query lengths of 7 to 204 tokens. The indexing and timing setup is described in Appendix~\ref{app:setup}.

\label{sec:eff-bert-overall}


\subsection{Retrieval Latency}
\label{sec:overall-mrt}
Table~\ref{tab:mrt} reports retrieval mean response time (MRT), decomposed into query-encoding and search time, for both BERT and ModernBERT backbones.
Here, unless otherwise stated, FLINT uses $V{=}32$.
On the three long-query corpora, FLINT achieves
$1.6$--$2.8\times$ faster retrieval than ConstBERT with BERT
and $1.5$--$2.7\times$ with ModernBERT.
The largest gains occur on TREC-ToT 2025, where MRT falls
from 74.3 to 26.1\,ms with BERT and from 140.9 to 51.3\,ms
with ModernBERT.
Relative to ColBERTv2, FLINT is $10.5$--$36.2\times$ faster
with BERT and $23$--$170\times$ faster with ModernBERT
on these corpora.
On NFCorpus, whose queries average only seven tokens,
FLINT and ConstBERT have very similar MRT.
\textbf{FLINT's latency advantage over the late-interaction
baselines is strongest on long queries.}

Figure~\ref{fig:mrt-breakdown} illustrates where these latency gains
arise, whether from query encoding or search.
As shown in the figure, FLINT and ConstBERT differ in how
their MRT is distributed between encoding and search.
With the same backbone, FLINT and ConstBERT have similar
query-encoding costs.
Their search times, however, diverge sharply as search time accounts for most of ConstBERT's MRT on the
long-query corpora. For example,
FLINT searches $2.5$--$4.6\times$ faster with BERT and
$2.5$--$4.8\times$ faster with ModernBERT.
\textbf{FLINT's lower MRT comes primarily from faster search,
while its query-encoding time remains similar to that of
the baselines.}

The ModernBERT rows of Table~\ref{tab:mrt} extend this comparison to the
single-vector baseline and FLINT with $V \in \{16, 32, 64\}$.
Single-Vector is the fastest ModernBERT system, since its search takes about 1\,ms, but it is far less effective (Section~\ref{sec:modernbert-backbone}).

Section~\ref{sec:budget} showed how the budget $V$ affects
effectiveness; here, we report its latency cost.
Compared with the single-vector baseline, FLINT with $V{=}16$
adds 6--16\,ms of MRT while improving effectiveness across
all benchmarks.
Increasing $V$ further raises search cost. For example,
$V{=}16$ searches in at most 12.1\,ms and $V{=}32$ in at most
23.6\,ms, whereas $V{=}64$ takes 31.4--75.1\,ms on the
long-query corpora, $2.8$--$4.9\times$ the search time
of $V{=}32$.
\textbf{FLINT's budget provides a controllable trade-off
between effectiveness and latency, with the additional
effectiveness gains from larger budgets coming at higher
search cost.}

\subsection{Latency Scaling with Query Length}
\label{sec:eff-qlen}

FLINT's latency advantage increases with query length because it always
produces a constant number of $V$ (here $V$=32) query vectors and thus its search cost remains nearly constant, whereas ConstBERT retains one representation per query token and therefore
becomes increasingly expensive on longer queries.
As shown in Figure~\ref{fig:mrt-decompose},  on TREC-ToT 2025 with BERT, the encoding time for longer queries increases at the same rate for both  FLINT and ConstBERT models (left). However, their search times diverge sharply (middle). FLINT's search time remains nearly constant relative to number of vectors in the query, with a fitted slope of $0.004$\,ms per token, compared with $0.31$\,ms for ConstBERT. Consequently, FLINT's total MRT grows only slightly with query length, while ConstBERT's increases much more steeply (right).
 The same pattern holds with ModernBERT. Truncating ConstBERT's queries to 32 tokens would also cap its query-vector count, but reduces average NDCG@10 on every complex-query suite with both backbones (Appendix~\ref{app:trunc}).

\begin{figure}[t]
  \centering
  \vspace{-2em}
  \captionsetup{}
  \includegraphics[width=\linewidth]{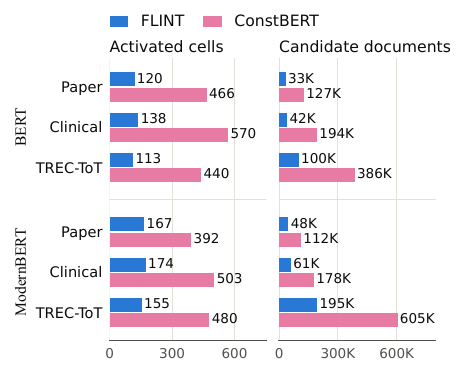}
  \caption{Mean activated PLAID cells and candidate documents per query for
  FLINT and ConstBERT on long-query corpora.}
  \label{fig:plaid-cells}
\end{figure}

\subsection{PLAID Cell Activation}
\label{sec:eff-topology}
PLAID does not score every document in the corpus. Each query vector probes
its nearest centroids, activating the corresponding cells and contributing documents to the candidate pool. The union of the corresponding cells determines the candidate documents that are then scored with MaxSim.
Therefore, {activating more cells broadens the search and generally increases the number of candidates; activating fewer cells reduces search work.} However, fewer activated cells are useful only insofar as the resulting candidate pool is still effective.
Figure~\ref{fig:plaid-cells} shows that FLINT consistently activates fewer cells than ConstBERT on the three long-query corpora. This translates into fewer candidate documents, as ConstBERT scores $2.3$--$4.6\times$ more candidates across the two backbones. It also compares each candidate against $5$--$7\times$ more query vectors. \textbf{FLINT therefore reduces search work by scoring fewer candidates and performing fewer vector comparisons per candidate.}

The reduction in activated cells is larger on BERT, where FLINT's pooled vectors are also more concentrated in direction and probe fewer distinct cells (Appendix~\ref{app:effrank}). These savings accompany competitive effectiveness relative to ConstBERT on conventional retrieval and stronger results on the complex-query suites, showing that FLINT can search a smaller candidate pool without sacrificing overall retrieval effectiveness.

\begin{table}[h]
\vspace{1em}
  \centering\footnotesize
  \setlength{\tabcolsep}{2.5pt}
  \renewcommand{\arraystretch}{1.0}
  \adjustbox{max width=\linewidth}{%
  \begin{tabular}{@{}ll|rrrr@{}}
  \toprule
   & \textbf{Architecture}
    & \makecell[c]{\textbf{NFCorpus}\\{\scriptsize 3.6K docs}}
    & \makecell[c]{\textbf{CRUMB}\\\textbf{Paper}\\{\scriptsize 363K docs}}
    & \makecell[c]{\textbf{CRUMB}\\\textbf{Clinical}\\{\scriptsize 915K docs}}
    & \makecell[c]{\textbf{TREC-ToT}\\\textbf{2025}\\{\scriptsize 6.4M docs}} \\
  \midrule
  \multirow{3}{*}{\rotatebox[origin=c]{90}{BERT}} & FLINT ($V{=}32$) & \underline{5.3M} & 440M & 1.1G & 7.3G \\
   & ConstBERT & \underline{5.3M} & 444M & 1.1G & 7.4G \\
   & ColBERTv2 & 29M  & 2.5G & 6.4G & 39G \\
  \midrule
  \multirow{6}{*}{\rotatebox[origin=c]{90}{ModernBERT}} & Single-Vector & 5.5M          & \textbf{38M} & \textbf{84M} & \textbf{477M} \\
   & FLINT ($V{=}16$) & \textbf{2.7M} & \underline{228M} & \underline{563M}  & \underline{3.8G} \\
   & FLINT ($V{=}32$) & 5.4M          & 444M & 1.1G  & 7.5G \\
   & FLINT ($V{=}64$) & 9.6M          & 885M & 2.2G  & 15G \\
   & ConstBERT & \underline{5.3M}          & 440M & 1.1G  & 7.5G \\
   & ColBERTv2 & 41M           & 3.0G & 10.0G & 70G \\
  \bottomrule
  \end{tabular}%
  }
  \caption{On-disk index size per corpus ($\downarrow$ lower is better).
  \textbf{Bold} marks the smallest index per corpus and \underline{underline} the second smallest.}
  \label{tab:index-size}
\end{table}

\subsection{Index Size}
\label{sec:corpus-storage}

Table~\ref{tab:index-size} reports on-disk index sizes. Because FLINT and
ConstBERT both store 32 vectors per document, their BERT indices are nearly identical, so \textbf{FLINT's latency reductions do not come at the cost of a larger index.}
 The same holds on ModernBERT. Compared with token-level ColBERTv2, FLINT's indices are $5.3$--$5.8\times$ smaller with BERT and $6.8$--$9.3\times$ smaller with ModernBERT. 

The fixed document-vector budget makes storage per document independent of document length, with index size scaling with $V$ instead. This provides direct control over storage requirements. Single-vector retrieval yields even smaller indices on the three larger corpora, but at a substantial cost in effectiveness (Section~\ref{sec:effectiveness}). 
These results show that \textbf{FLINT's fixed document-vector budget prevents the index from growing with document length.} Combined with the response-time
results above, FLINT provides the storage efficiency of ConstBERT while
achieving markedly lower latency than both ConstBERT and token-level
retrieval on the larger, long-query corpora.

\section{Conclusion}
\label{sec:conclusion}

We introduced FLINT, a fixed-budget late-interaction architecture that addresses two central scaling challenges: document storage and query-side search cost. Shared attention-based pooling maps both queries and documents to a fixed number of vectors, preserving multi-vector matching without retaining one vector per token.

FLINT achieves $5.3$--$9.3\times$ smaller indices than token-level models and, on the long-query corpora, $1.5$--$2.8\times$ faster end-to-end retrieval than ConstBERT. Its search latency remains nearly constant as queries grow longer. With the vector budget $V$ controlling the effectiveness--efficiency trade-off, FLINT makes storage and search computation explicit architectural choices, supporting efficient retrieval for long and complex queries.

\section*{Limitations}
\label{sec:limitations}

\paragraph{Training distribution.} Our BERT FLINT model is
supervised on the same MS MARCO passage triples as ColBERTv2 and ConstBERT.
We adopt this deliberately, so that differences between FLINT, ConstBERT, and
ColBERTv2 reflect the pooling architecture rather than the training data, but it
means the only query distribution the pooling module ever observes during
training is MS MARCO's short, keyword-like queries. Our central claims concern
long, complex queries, so those results are entirely zero-shot generalization. To our knowledge, there is no comparable MS MARCO-scale dataset of high-quality long queries. Moreover, multi-vector retrievers are known to be sensitive to both the scale and quality of training data. We therefore leave long-query-specific training to future work.

\paragraph{Limited budget sweep.} On BERT we fix the budget at $V{=}32$,
matching ConstBERT's setting so the two fixed-budget systems are directly
comparable. We vary $V \in \{16, 32, 64\}$ only on ModernBERT, so it remains
open whether symmetric pooling degrades more or less gracefully than
document-only pooling at smaller budgets, and whether a larger budget would
close FLINT's remaining gap to ColBERTv2 on BEIR. We also tie the query and
document budgets to the same value. Decoupling them, for instance a smaller
query budget paired with a larger document budget, is unexplored.

\paragraph{Scope of the evaluation.} On BERT, absolute effectiveness on the
complex-query suites is low for every system we evaluate, with the best NDCG@10 on TREC Tip-of-the-Tongue 2025 reaching only $0.065$. FLINT's
improvements in this regime are consistent across suites, but they are
improvements within a band where all late-interaction models we test perform
poorly, and the small absolute margins should be read with that in mind. Our
evaluation also covers two encoder-only backbones of modest scale, so behavior
under substantially larger backbones remains untested.


\section*{GenAI Usage Disclosure}
We used large language models to edit the grammar and wording of this paper,
to assist with coding, and to help clean up the
code repository for release. The research design, analysis, interpretation of
results, and conclusions are entirely our own.


\bibliography{custom}
\appendix

\section{Related Work}
\label{app:related}

\paragraph{Fixed-size multi-vector representations.}
MetaEmbed \citep{xiao2025metaembed} appends learnable tokens to the input of a
vision-language model and uses their final hidden states as multi-vector
embeddings for multimodal retrieval, trained with Matryoshka-style nested
groups so the number of vectors can be chosen at test time. FLINT instead targets long-query text
retrieval and pools the encoder outputs with a separate cross-attention module
rather than adding tokens to the input. It uses a fixed budget $V$ per model
instead of nested groups, and projects each vector to a low dimension, whereas
MetaEmbed keeps the full hidden dimension, creating much larger indices. They also focus primarily on multi-modal benchmarks rather than on long and complex textual datasets.

Poly-encoders \citep{Humeau2020Poly-encoders} use learned codes that attend over
the query tokens to produce several query vectors, while each document remains a
single vector compared through attention. FLINT pools both queries and documents
into multiple vectors and scores them with MaxSim.

ConstBERT \citep{constbert} pools each document into a fixed number of vectors
with a learned linear layer and keeps queries token-level without compression. FLINT pools
both sides with cross-attention, so query cost does not grow with query length.

\paragraph{Reducing token vectors.}
Token pooling \citep{clavie2024reducing} clusters document token vectors post-hoc, and static pruning
\citep{acquavia2023static} removes low-impact document vectors. As presented in both papers, however, the
number of vectors still grows with document length and queries are unchanged,
whereas FLINT learns a fixed number of vectors for both sides.

Query embedding pruning \citep{tonellotto2021qep} keeps the query token vectors
of the rarest terms for candidate generation. FLINT instead learns to summarize
all query tokens into $V$ vectors and also fixes the document side vector count.

\paragraph{Efficient multi-vector search.}
XTR \citep{lee2023xtr} trains the model so that its first-stage token retrieval
finds the relevant document tokens, then scores documents using only those
retrieved tokens. This skips loading every candidate's full set of vectors, but
it keeps one vector per token for queries and documents, so its index is as
large as ColBERT's. WARP \citep{scheerer2025warp} is a faster search engine
for XTR-style retrieval, but its index still stores one vector per token. MUVERA \citep{dhulipala2024muvera} is a latency-focused
prefilter: it compresses each multi-vector set into one fixed-dimensional
encoding (FDE) for fast single-vector candidate search, then reranks candidates
with the full token-level vectors. Because the FDE index sits on top of the full
multi-vector index, its storage exceeds that of the multi-vector index alone, whereas FLINT reduces
storage.

\paragraph{Single-vector retrieval.}
With $V{=}32$ vectors of dimension 128, FLINT represents each input with 4096
values. Single-vector models that produce vector embeddings with the same amount of floats, such as RepLLaMA
\citep{ma2024fine}, rely on much larger backbones: it fine-tunes LLaMA-7B as a
bi-encoder producing one 4096-dimensional vector scored by dot product. To our
knowledge, no single-vector model with a similarly sized and performing backbone produces
embeddings of this size. A size-matched comparison would thus alter the backbone's strength, and we therefore leave it out of scope. We instead compare against a single-vector model with the same backbone and
training data (Section~\ref{sec:modernbert-backbone}).

\paragraph{Other approaches.}
TRIAL \citep{kang-etal-2025-trial} improves the scoring function rather than compressing representations; SCV \citep{park-etal-2025-scv} compresses only documents via span pooling with a single-vector prefilter, so query-side cost still scales with query length; Zong \& Piwowarski \citep{zong2025towards} dynamically prune document tokens, with pruning ratio varying per document and query tokens left uncompressed. FLINT compresses queries and documents to a fixed, input-length-independent budget.

\section{Experimental Settings}
\label{app:setup}

\textbf{Training.} On BERT, FLINT is trained on MS MARCO passage-ranking
triples with 64 candidates per query, in fp16 with a global batch size of 48,
a learning rate of $10^{-5}$, and a linear schedule over 267K steps. On ModernBERT-large, we train the FLINT
($V \in \{16, 32, 64\}$), ConstBERT ($32$ document vectors), and ColBERTv2
architectures ourselves (ModernBERT FLINT, ModernBERT ConstBERT, and ModernBERT
ColBERTv2) under the multi-vector training regimen of
\citet{warner2024modernbert}, on MS MARCO with 32 candidates per query, for one
epoch in bf16 with a global batch size of 16, a learning rate of
$3 \times 10^{-5}$, and 5\% warmup. The FLINT pooling module uses a learning
rate of $10^{-4}$.
Training queries are truncated to 84 tokens. The single-vector baseline is a
publicly available ModernBERT-large
bi-encoder,\footnote{\url{https://huggingface.co/joe32140/ModernBERT-large-msmarco}}
trained identically to the single-vector retriever introduced in \citep{warner2024modernbert}.

\textbf{Pooling module.} The pooling module uses $H{=}8$ attention heads and a
feed-forward expansion ratio of $r{=}4$ on BERT ($h{=}768$) and $r{=}2$ on
ModernBERT ($h{=}1024$), with dropout 0.1. Pooled vectors are projected to $h'{=}128$
dimensions. On BERT, the query and document sides use separate projections; on
ModernBERT, they share one.

\textbf{Sequence lengths.} At evaluation, queries are truncated to $M{=}512$
tokens for every model. The evaluation document length $N$ differs between backbones because of ConstBERT:
its learned pooling maps a fixed number of token positions to its 32 document
vectors, so $N$ is fixed when ConstBERT is trained. The released BERT ConstBERT
checkpoint uses $N{=}250$, so we truncate documents to $N{=}250$ for all BERT
models. On ModernBERT, we train ConstBERT with $N{=}512$ and truncate documents
to $N{=}512$ for all ModernBERT models. Within each backbone, every model
therefore sees the same document context.

\textbf{Query padding.} BERT ConstBERT and ColBERTv2 pad queries shorter than
32 tokens up to 32, matching ColBERTv2's default and the setting they were
trained with. All ModernBERT models are trained without query padding and are
evaluated the same way. BERT FLINT queries are not padded during training or
evaluation, and FLINT always produces $V$ query vectors.

\textbf{Indexing and retrieval.} Multi-vector models are indexed with PLAID
\citep{santhanam2022plaid,macavaney2024plaid} using 2-bit residual
compression. For the efficiency results, every
multi-vector model uses PLAID with \texttt{ncells}$=8$ and $k=1000$ retrieved
candidates, and the single-vector model uses a FAISS OPQ-IVFPQ index
\citep{5432202}.

\textbf{Latency measurement.} All latency experiments run on a single NVIDIA
A100 80GB GPU. All systems for a corpus are timed on the same machine. After a global warm-up, we take the median of three timed
runs per query. Query-encoding time
covers the model forward pass, and search time covers candidate retrieval and
scoring.

\section{Dataset Details}
\label{app:datasets}

Table~\ref{tab:datasets} provides detailed statistics for the four evaluation
suites, including corpus and query counts, mean document and query lengths,
task focus, and query characteristics. Lengths are measured using the ColBERT
tokenizer. For suites containing multiple corpora or tasks, we report the
range across the evaluated collections. We exclude CRUMB Legal and SetOps from our tables since queries are short (148 and 47 characters, respectively vs. 415 for Theorem, the next shortest). They did not fit in our curation of long query benchmarks. We report their results, along with MS MARCO, in Table~\ref{tab:extra-results}.

\section{Query Geometry and Effective Rank}
\label{app:effrank}

Section~\ref{sec:eff-topology} shows that FLINT activates fewer PLAID cells
than ConstBERT. Beyond the number of query vectors, the cells a query
activates also depend on how spread out its vectors are: vectors pointing in
similar directions tend to probe the same centroids. We quantify this spread
using effective rank, or the participation ratio \citep{gao2017theory}:
\begin{equation}
\mathrm{erank}(Q)=
\frac{\left(\sum_i \lambda_i\right)^2}
     {\sum_i \lambda_i^2},
\qquad
\lambda_i=\sigma_i^2(Q),
\end{equation}
where lower effective rank indicates that the query vectors are concentrated
along fewer directions.

\begin{figure*}[t]
  \centering
  \captionsetup{}
  \includegraphics[width=0.8\textwidth]{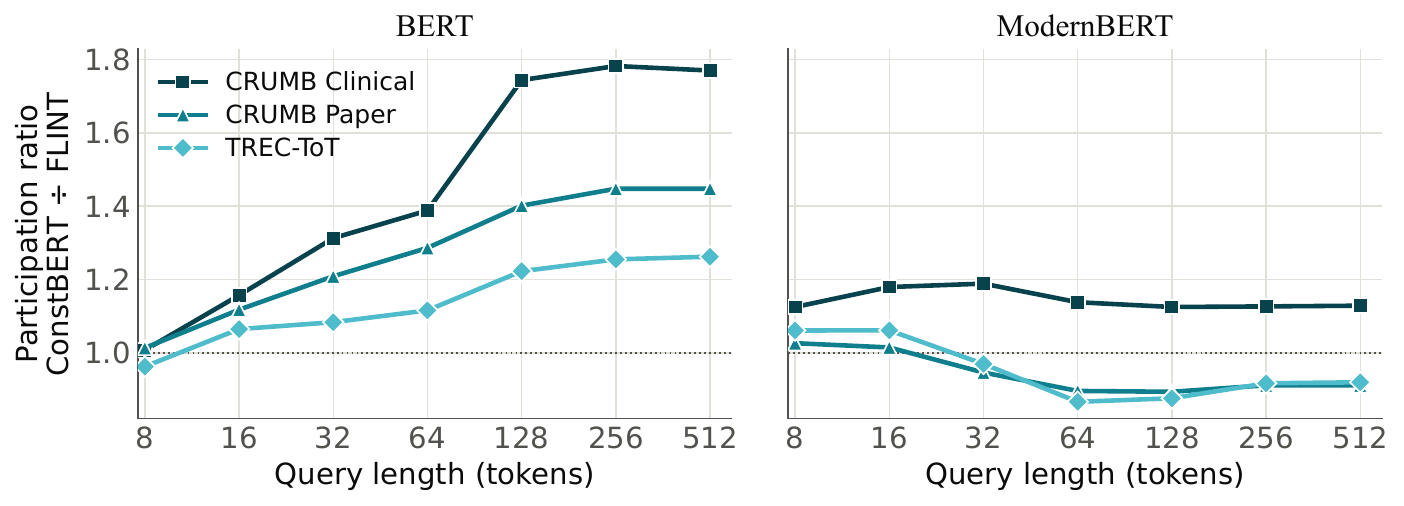}
  \caption{Ratio of ConstBERT's to FLINT's query-vector participation
  ratio vs.\ query length, one line per corpus. Above the dotted line,
  ConstBERT's query vectors are more spread out than FLINT's.}
  \label{fig:effrank-length}
\end{figure*}

\textbf{Spread of the query vectors.} Figure~\ref{fig:effrank-length} divides
ConstBERT's participation ratio by FLINT's at each query length. On BERT, the two
are nearly equal on the shortest queries ($0.96$--$1.01$ at 8 tokens), and
ConstBERT's ratio grows faster as queries get longer, reaching $1.78\times$
FLINT's on CRUMB Clinical, $1.45\times$ on CRUMB Paper, and $1.26\times$ on
TREC-ToT 2025. FLINT's pooled vectors therefore stay more concentrated than
ConstBERT's token vectors. On ModernBERT, the ratio stays between $0.87$ and
$1.19$ at every length, so the two models' query vectors are about equally
spread out.

\begin{figure*}[t]
  \centering
  \captionsetup{}
  \vspace{-1.5em}
  \includegraphics[width=\textwidth]{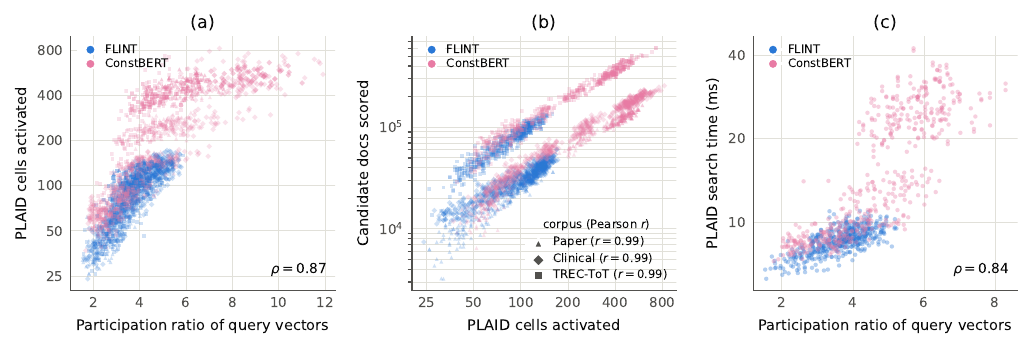}
  \caption{Query spread and PLAID search on BERT.
  (a) Participation ratio vs.\ PLAID cells activated.
  (b) Cells activated vs.\ candidate documents scored.
 (c) Participation ratio vs.\ PLAID search time (CRUMB Paper).
  One point per query ($n{=}3{,}248$ in (a)--(b), $1{,}008$ in (c));
  all axes are log except participation ratio. $\rho$ is pooled Spearman, $r$ is
  within-corpus Pearson.}
  \label{fig:effrank-mechanism}
\end{figure*}

\textbf{Spread and cell activation on BERT.} Figure~\ref{fig:effrank-mechanism}
connects this spread to PLAID search on BERT. Across $3{,}248$ queries, the
participation ratio is strongly associated with the number of activated cells
($\rho=0.87$), and cell activation closely tracks the number of candidate
documents scored (within-corpus $r=0.99$). The participation ratio also
correlates with PLAID-search latency on CRUMB Paper ($\rho=0.84$,
$n=1{,}008$), although this pooled correlation also reflects ConstBERT's query
length.

\begin{figure*}[t]
  \centering
  \captionsetup{}
  \includegraphics[width=0.8\textwidth]{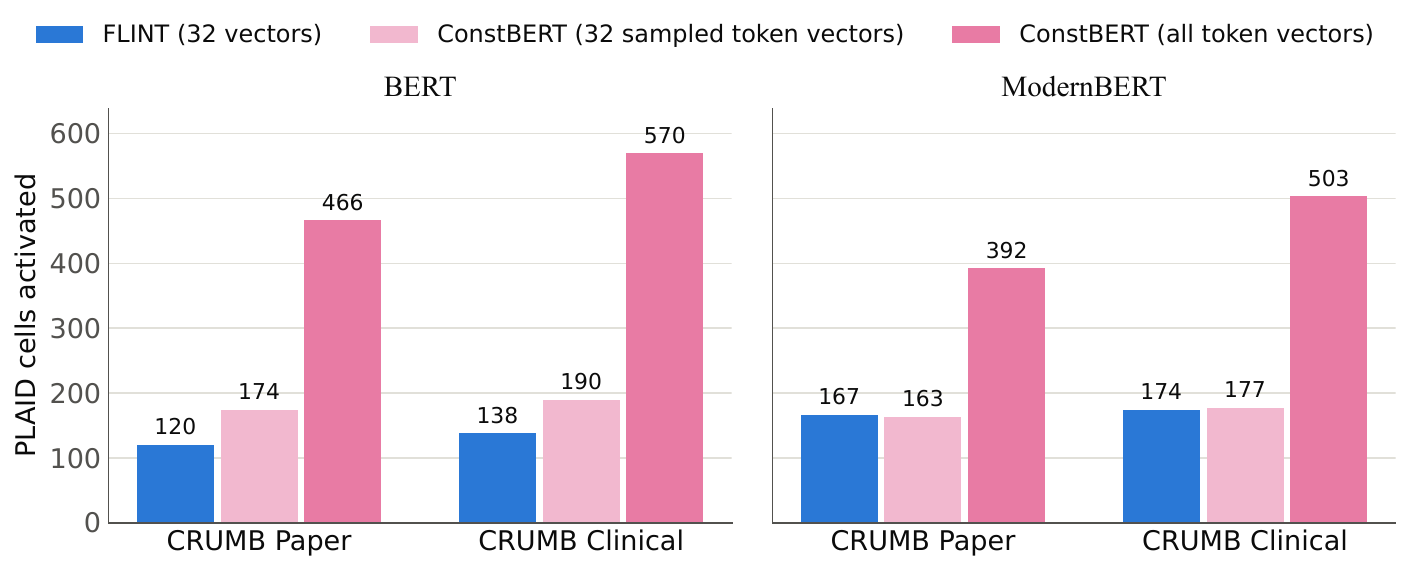}
  \caption{PLAID cells activated at an equal number of query vectors.
  FLINT's 32 pooled vectors compared with 32 of ConstBERT's token vectors
  sampled at random per query, and with all of ConstBERT's token vectors.}
  \label{fig:cells-vs-vectors}
\end{figure*}

\textbf{Cells at the same number of query vectors.} Both the number of query
vectors and their spread affect how many cells a query activates. To separate
the two, we give ConstBERT the same number of vectors as FLINT by randomly
sampling 32 of its token vectors per query (Figure~\ref{fig:cells-vs-vectors}).
On BERT, FLINT's 32 vectors still activate fewer cells than ConstBERT's 32
(120 vs.\ 174 on CRUMB Paper, 138 vs.\ 190 on CRUMB Clinical), which matches
their lower spread. On ModernBERT, the two activate the same number of cells
(167 vs.\ 163 and 174 vs.\ 177), which matches their equal spread. With all of
its token vectors, ConstBERT activates $2.3$--$4.1\times$ more cells than FLINT
on both backbones. Most of FLINT's reduction in activated cells therefore comes
from using fewer query vectors, and on BERT the lower spread of its pooled
vectors adds to it.

\textbf{Pooling variants.} Table~\ref{tab:effrank-stats} reports mean query and
document effective ranks for ColBERTv2, ConstBERT, and FLINT-symmetric, and for
BERT-FLINT architecture variants. Primary FLINT-symmetric shares the perceiver
across both sides; unless marked query-side, the variant block changes only the
\emph{document-side} pooler. Deep and iterative-shallow doc-side variants sit
near $\mathrm{erank}\!\approx\!1$, indicating near-total directional collapse
of the latent set, whereas no-SA and the query-side perceiver retain higher
ranks closer to ConstBERT and ColBERTv2. FLINT-symmetric sits below ConstBERT
on query rank, consistent with the BERT results above.

\section{Recall Statistics for Complex Query Benchmarks}

Table~\ref{tab:complex-recall} reports Recall@1000 on the three
complex-query suites, complementing the NDCG@10 results of
Table~\ref{tab:ndcg-results}. With the BERT backbone, FLINT recalls
more than ConstBERT on Clinical ($0.351$ vs.\ $0.306$) and
TREC-ToT ($0.323$ vs.\ $0.277$), while ConstBERT recalls more
on Paper ($0.443$ vs.\ $0.400$) and Theorem ($0.382$ vs.\ $0.246$); the two
models are close on CRUMB overall ($0.372$ vs.\ $0.395$) and on BrowseComp+
($0.391$ vs.\ $0.401$). With ModernBERT, FLINT ($V{=}32$) recalls more than
ConstBERT on every collection, with the largest gaps on TREC-ToT ($0.458$ vs.\
$0.272$) and BrowseComp+ ($0.555$ vs.\ $0.466$). It also recalls more than
ModernBERT ColBERTv2 on every collection except CRUMB ToT ($0.526$ vs.\
$0.578$), including CRUMB overall ($0.508$ vs.\ $0.441$), TREC-ToT ($0.458$
vs.\ $0.402$), and BrowseComp+ ($0.555$ vs.\ $0.526$). Raising the budget to
$V{=}64$ further improves recall (CRUMB $0.525$, BrowseComp+ $0.576$), while
the single-vector model trails every FLINT variant on CRUMB and BrowseComp+.

\section{Query Truncation for ConstBERT}
\label{app:trunc}

Table~\ref{tab:constbert-trunc} reports ConstBERT's effectiveness when its
queries are truncated to 32 tokens, which gives it at most as many query
vectors as FLINT. Truncation lowers average NDCG@10 on CRUMB from $0.136$ to
$0.095$, on TREC-ToT 2025 from $0.047$ to $0.018$, and on BrowseComp+ from
$0.055$ to $0.037$. StackExchange is the only collection where it helps
($0.170$ vs.\ $0.140$). ModernBERT ConstBERT
shows the same pattern: truncation lowers average NDCG@10 from $0.180$ to
$0.127$ on CRUMB, from $0.070$ to $0.032$ on TREC-ToT 2025, and from $0.067$ to
$0.036$ on BrowseComp+, and StackExchange is essentially unchanged ($0.209$ vs.\
$0.207$). Truncating the query is therefore very lossy on both backbones, and it
does not substitute for learned compression of the query into a fixed budget.

\begin{table}[h]
  \centering\footnotesize
  \setlength{\tabcolsep}{4pt}
  \renewcommand{\arraystretch}{1.0}
  \begin{tabular}{@{}ll|ccc@{}}
  \toprule
  & & \textbf{MS MARCO} & \multicolumn{2}{c}{\textbf{CRUMB}} \\
  \cmidrule(lr){3-3}\cmidrule(l){4-5}
   & \textbf{Architecture} & MRR@10 & Legal & SetOps \\
  \midrule
  \multirow{3}{*}{\rotatebox[origin=c]{90}{BERT}} & FLINT ($V{=}32$) & 0.380 & 0.138 & 0.184 \\
   & ConstBERT & 0.387 & 0.136 & 0.232 \\
   & ColBERTv2 & \underline{0.395} & 0.158 & \textbf{0.290} \\
  \midrule
  \multirow{6}{*}{\rotatebox[origin=c]{90}{ModernBERT}} & Single-Vector & 0.363 & 0.138 & 0.044 \\
   & FLINT ($V{=}16$) & \underline{0.395} & 0.154 & 0.177 \\
   & FLINT ($V{=}32$) & \textbf{0.405} & \underline{0.175} & 0.208 \\
   & FLINT ($V{=}64$) & \textbf{0.405} & 0.159 & 0.227 \\
   & ConstBERT & 0.386 & \textbf{0.182} & 0.187 \\
   & ColBERTv2 & 0.366 & 0.123 & \underline{0.252} \\
  \bottomrule
  \end{tabular}
  \caption{Results on MS MARCO (dev, MRR@10) and the CRUMB Legal and SetOps
  tasks (NDCG@10), which are not included in Table~\ref{tab:ndcg-results}.
  \textbf{Bold} indicates the best result in each column and
  \underline{underline} the second best.}
  \label{tab:extra-results}
\end{table}


\vspace{0.6em}

\begin{table*}[b]
  \centering\footnotesize
  \captionsetup{font=footnotesize}
  \setlength{\tabcolsep}{5pt}
  \renewcommand{\arraystretch}{1.1}
  \begin{tabular}{@{}llcccc@{}}
  \toprule
  & & \multicolumn{2}{c}{\textbf{CRUMB Paper}}
    & \multicolumn{2}{c}{\textbf{TREC-ToT}} \\
  \cmidrule(lr){3-4}\cmidrule(lr){5-6}
  \textbf{Architecture} & \textbf{Pooling}
  & $Q_{\mathrm{eff}}$ & $D_{\mathrm{eff}}$
  & $Q_{\mathrm{eff}}$ & $D_{\mathrm{eff}}$ \\
  \midrule
  \multicolumn{6}{@{}l}{\textit{Reference systems}} \\
  ColBERTv2
    & token-level
    & 4.28 & 4.17 & 3.92 & 4.02 \\
  ConstBERT
    & doc-side only
    & 4.24 & 3.52 & 3.54 & 2.75 \\
  FLINT-symmetric (primary)
    & both sides
    & 3.54 & 3.83 & 3.23 & 3.72 \\
  \midrule
  \multicolumn{6}{@{}l}{\textit{BERT-FLINT variants}} \\
  FLINT, no\,SA
    & doc-side
    & 3.91 & 3.69 & 3.53 & 3.35 \\
  FLINT (1\,SA)
    & doc-side
    & 2.71 & 1.79 & 2.50 & 1.93 \\
  FLINT, query-side perceiver
    & query-side
    & 5.28 & 3.58 & 4.66 & 4.06 \\
  Deep FLINT (1CA{+}6SA)
    & doc-side
    & 1.55 & 1.07 & 1.44 & 1.08 \\
  Iterative shallow FLINT $(1\mathrm{CA}{+}2\mathrm{SA}){\times}3$
    & doc-side
    & 1.53 & 1.08 & 1.47 & 1.09 \\
  \bottomrule
  \end{tabular}
  \caption{Mean query and document effective rank.
  Participation-ratio effective ranks averaged over a fixed sample of
  5{,}000 documents and all available queries (CRUMB Paper: 72; TREC-ToT
  train: 143).
  Lower values indicate more redundant multi-vector sets.
  FLINT-symmetric applies the shared perceiver to both sides
  (primary configuration).
  Variant rows below are \emph{document-side} pooling changes unless marked
  query-side; CA\,=\,cross-attention and SA\,=\,self-attention in the pooling
  module; $(1\mathrm{CA}{+}2\mathrm{SA}){\times}3$ denotes three stacked
  (1\,CA{+}2\,SA) stages.}
  \label{tab:effrank-stats}
\end{table*}

\begin{table*}[b]
  \centering
  \footnotesize
  \captionsetup{font=footnotesize}
  \setlength{\tabcolsep}{6pt}
  \renewcommand{\arraystretch}{1.05}

  \adjustbox{max width=\linewidth}{%
  \begin{tabular}{ll|ccccccc|c|c}
    \toprule
    \multicolumn{2}{c|}{}
    & \multicolumn{7}{c|}{\textbf{CRUMB}}
    & \multicolumn{1}{c|}{\textbf{TREC ToT 2025}}
    & \multicolumn{1}{c}{\textbf{BrowseComp+}} \\
    \cmidrule(lr){3-9}
    \cmidrule(lr){10-10}
    \cmidrule(lr){11-11}

    \textbf{Backbone}
    & \textbf{Architecture}
    & \shortstack{Code\\{\scriptsize $\bar q$:\,229}}
    & \shortstack{StackEx\\{\scriptsize $\bar q$:\,226}}
    & \shortstack{Clinical\\{\scriptsize $\bar q$:\,183}}
    & \shortstack{Paper\\{\scriptsize $\bar q$:\,180}}
    & \shortstack{ToT\\{\scriptsize $\bar q$:\,168}}
    & \shortstack{Theorem\\{\scriptsize $\bar q$:\,97}}
    & \shortstack{\textbf{Avg}\\{\scriptsize $\bar q$:\,180}}
    & \shortstack{\textbf{Avg}\\{\scriptsize $\bar q$:\,204}}
    & \shortstack{\textbf{Avg}\\{\scriptsize $\bar q$:\,117}} \\
    \midrule

    \multicolumn{2}{l}{}
    & \multicolumn{9}{c}{\textbf{Recall@1000}} \\

    BERT & FLINT
    & 0.093 & 0.756 & 0.351 & 0.400 & 0.385 & 0.246 & 0.372 & 0.323 & 0.391 \\

    BERT & ConstBERT
    & 0.096 & 0.797 & 0.306 & 0.443 & 0.348 & 0.382 & 0.395 & 0.277 & 0.401 \\

    BERT & ColBERTv2
    & 0.070 & 0.791 & 0.277 & 0.520 & 0.437 & 0.409 & 0.417 & 0.348 & 0.407 \\

    \addlinespace[3pt]

    \rowcolor{gray!12}
    ModernBERT & Single-Vector
    & 0.085 & 0.820 & 0.358 & 0.409 & 0.326 & 0.340 & 0.389 & 0.327 & 0.339 \\

    \rowcolor{gray!12}
    ModernBERT & FLINT ($V{=}16$)
    & 0.161 & 0.908 & \textbf{0.451} & 0.479 & 0.467 & 0.370 & 0.472 & 0.312 & 0.502 \\

    \rowcolor{gray!12}
    ModernBERT & FLINT ($V{=}32$)
    & 0.193 & 0.915 & 0.428 & \textbf{0.533} & 0.526 & 0.451 & 0.508 & 0.458 & 0.555 \\

    \rowcolor{gray!12}
    ModernBERT & FLINT ($V{=}64$)
    & \textbf{0.199} & \textbf{0.930} & 0.440 & 0.532 & 0.556 & \textbf{0.494} & \textbf{0.525} & \textbf{0.529} & \textbf{0.576} \\

    \rowcolor{gray!12}
    ModernBERT & ConstBERT
    & 0.155 & 0.909 & 0.386 & 0.382 & 0.422 & 0.443 & 0.449 & 0.272 & 0.466 \\

    \rowcolor{gray!12}
    ModernBERT & ColBERTv2
    & 0.137 & 0.889 & 0.228 & 0.416 & \textbf{0.578} & 0.397 & 0.441 & 0.402 & 0.526 \\

    \bottomrule
  \end{tabular}%
  }

  \caption{Retrieval coverage on the complex-query benchmarks,
  measured using Recall@1000.
  {\setlength{\fboxsep}{1pt}\colorbox{gray!12}{Shaded rows}} use
  ModernBERT, while unshaded rows use BERT.
  $V$ is the number of fixed vectors per query/document emitted by the
  FLINT perceiver.
  \textbf{Bold} indicates the best result.}
  \label{tab:complex-recall}
\end{table*}

\begin{table*}[b]
  \centering
  \footnotesize
  \captionsetup{font=footnotesize}
  \setlength{\tabcolsep}{6pt}
  \renewcommand{\arraystretch}{1.05}

  \adjustbox{max width=\linewidth}{%
  \begin{tabular}{ll|ccccccc|c|c}
    \toprule
    \multicolumn{2}{c|}{}
    & \multicolumn{7}{c|}{\textbf{CRUMB}}
    & \multicolumn{1}{c|}{\textbf{TREC ToT 2025}}
    & \multicolumn{1}{c}{\textbf{BrowseComp+}} \\
    \cmidrule(lr){3-9}
    \cmidrule(lr){10-10}
    \cmidrule(lr){11-11}
    \textbf{Backbone} & \textbf{Query length}
    & Code & StackEx & Clinical & Paper & ToT & Theorem & \textbf{Avg}
    & \textbf{Avg} & \textbf{Avg} \\
    \midrule
    BERT & $M{=}32$
    & 0.027 & \textbf{0.170} & 0.167 & 0.166 & 0.025 & 0.015 & 0.095 & 0.018 & 0.037 \\
    BERT & $M{=}512$
    & \textbf{0.049} & 0.140 & \textbf{0.225} & \textbf{0.296} & \textbf{0.075} & \textbf{0.027} & \textbf{0.136} & \textbf{0.047} & \textbf{0.055} \\
    \addlinespace[3pt]
    \rowcolor{gray!12}
    ModernBERT & $M{=}32$
    & 0.059 & \textbf{0.209} & 0.191 & 0.220 & 0.055 & 0.025 & 0.127 & 0.032 & 0.036 \\
    \rowcolor{gray!12}
    ModernBERT & $M{=}512$
    & \textbf{0.088} & 0.207 & \textbf{0.293} & \textbf{0.330} & \textbf{0.102} & \textbf{0.061} & \textbf{0.180} & \textbf{0.070} & \textbf{0.067} \\
    \bottomrule
  \end{tabular}%
  }

  \caption{ConstBERT NDCG@10 with queries truncated to $M$ tokens.
  Queries longer than $M$ tokens are truncated, and on BERT queries shorter than $M$ are padded up to 32. $M{=}512$ is the setting used throughout the paper
  (Table~\ref{tab:ndcg-results}).
  {\setlength{\fboxsep}{1pt}\colorbox{gray!12}{Shaded rows}} use ModernBERT.
  \textbf{Bold} marks the better query length per backbone.}
  \label{tab:constbert-trunc}
\end{table*}

%

\begin{table*}[b]
  \centering
  \scriptsize
  \captionsetup{font=footnotesize}
  \setlength{\tabcolsep}{3.5pt}
  \renewcommand{\arraystretch}{1.15}

  \begin{tabular}{
    @{}
    >{\raggedright\arraybackslash}p{1.45cm}
    >{\raggedleft\arraybackslash}p{1.05cm}
    >{\raggedleft\arraybackslash}p{0.90cm}
    >{\raggedleft\arraybackslash}p{0.85cm}
    >{\raggedleft\arraybackslash}p{0.75cm}
    >{\raggedright\arraybackslash}p{4.15cm}
    >{\raggedright\arraybackslash}p{5.25cm}
    @{}
  }
  \toprule
  \textbf{Suite} &
  \textbf{D} &
  \textbf{Q} &
  $\boldsymbol{\bar d}$ &
  $\boldsymbol{\bar q}$ &
  \textbf{Task focus} &
  \textbf{Query characteristics} \\
  \midrule

  BEIR\newline{\scriptsize 12 corpora set} &
  3.6K--5.4M &
  49--10K &
  14--384 &
  7--27 &
  Zero-shot heterogeneous retrieval across tasks including QA,
  fact-checking, and biomedical search. &
  Mostly short, keyword-like queries that evaluate transfer across
  heterogeneous in-domain and out-of-domain collections. \\

  \rowcolor{gray!10}
  BrowseComp+ &
  100K &
  830 &
  3810 &
  117 &
  Deep-research web-agent evaluation with human-verified supporting
  evidence. &
  Elaborate, multi-step information needs with hard distractors;
  much of the difficulty arises from identifying relevant evidence
  within long documents. \\

  CRUMB\newline{\scriptsize 6 long task subset} &
  24K--1.1M &
  69--3.7K &
  180--575 &
  97--229 &
  Unified multi-task suite for multi-aspect and reasoning-intensive
  retrieval. &
  Complex natural-language queries containing multiple explicit and
  implicit constraints, domain-specific terminology, and limited
  lexical overlap with relevant documents. \\

  \rowcolor{gray!10}
  TREC-ToT 2025 &
  6.4M &
  1.6K &
  654 &
  204 &
  Known-item entity re-finding from vague natural-language
  descriptions. &
  Very long, conversational, and unconstrained queries that may contain
  false memories, relative comparisons, and minimal direct lexical
  overlap with the target entity. \\

  \bottomrule
  \end{tabular}
  \caption{Evaluation benchmarks details.
  D and Q denote the numbers of documents and queries, while
  $\bar d$ and $\bar q$ denote their mean lengths in ColBERT tokens.
  Ranges indicate the minimum and maximum values across each dataset.}
  \label{tab:datasets}
\end{table*}

\end{document}